\documentstyle{article}

\def\ba{\begin{array}}
\def\ea{\end{array}}
\def\be{\begin{equation}}
\def\ee{\end{equation}}
\def\l{\lambda}

\def\D{{\rm D }}
\def\e{{\eplsilon}}
\def\p{{\phi}}

\def\d{{\rm d}}
\def\R{{\bf R}}
\def\bea{\begin{eqnarray}}
\def\eea{\end{eqnarray}}
\def\s{\sum}

\def\sq{\sqrt}

\def\r{\rho}

\def\e{\epsilon}

\def\I{\rm {I\kern-.3em I}}
\def\C{\rm {I\kern-.520em C}}
\def\R{\rm {I\kern-.3em R}}
\def\CZ{\rm {Z\kern-.4em Z}}

\def\unit{\rm {1\kern-.4em 1}}

\def\P{\rm P}
\begin{document}
\begin{titlepage}
\vspace{-10mm}
\hfill{}

\vskip 30 mm
\leftline{ \Large \bf
            Uniqueness of the minimum of the free energy }
\leftline{ \Large \bf
            of the 2D Yang--Mills theory at large $N$}

\vskip 10 mm
\leftline{A. Aghamohammadi$^{a,b,}${\footnote {e-mail:
           mohamadi@netware2.ipm.ac.ir}}, M. Alimohammadi$^{a,c,}${\footnote
           {e-mail:alimohmd@netware2.ipm.ac.ir}},
           M. Khorrami$^{a,c,d,}${\footnote{e-mail:mamwad@netware2.ipm.ac.ir}}}

\vskip 10 mm
{\it
  \leftline{ $^a$ Institute for Studies in Theoretical Physics and
            Mathematics, P.O.Box 5531, Tehran 19395, Iran}
\vskip -.4mm
  \leftline{ $^b$ Department of Physics, Alzahra University,
             Tehran 19834, Iran. }
\vskip -.4mm
  \leftline{ $^c$ Department of Physics, Tehran University,
             North-Kargar Ave. Tehran, Iran. }
\vskip -.4mm
  \leftline{ $^d$ Institute for Advanced Studies in Basic Sciences,
             P.O.Box 159, Gava Zang, Zanjan 45195, Iran. }}
\vskip 10mm
\begin{abstract}
There has been some controversies at the large $N$ behaviour of the
2D Yang-Mills and chiral 2D Yang-Mills theories. To be more specific,
is there a one parameter family of minima of the free energy in the strong
region, or the minimum is unique. We show that there is a missed equation
which, added to the known equations, makes the minimum unique.
\end{abstract}
\vskip 10 mm
\end{titlepage}
Recently the large group limit of the 2D Yang--Mills theory (YM$_2$)
has become of much interest from different points of view. There is a string
formulation  of the problem. It has been shown that there is  string theory
description of pure two dimensional gauge theories with various gauge groups
[1-5].
The large-group phase
transitions are also  of the interesting features of YM$_2$.
In ref. 6, the large $N$ behaviour of YM$_2$ for U($N$) gauge group on
the sphere has been studied. There, the free energy of YM$_2$
has been obtained for small areas (weak region).
Douglas and Kazakov (DK) \cite{DK} have studied this theory for arbitrary
areas and have shown that the theory has a third order phase transition at
some critical area ($A_c$). In their work, a symmetric ansatz for the
density function $\r $ is considered. In a subsequent work, Minahan and
Polychronakos (MP)\cite{MP} consider an asymmetric ansatz for the density
function and introduce a one parameter family of solutions.
They introduce a free parameter $Q$ (U(1) charge sector of U($N$)) and
express their results in terms of this parameter.

It was shown that the free energy of the complete SU($N$) theory is equal to
the $Q=0$ U(N) theory in the large area region, where the string expansion
is valid \cite{Ta}. In the words of  Crescimanno and Taylor (CT) \cite{CT},
the results of DK may be the complete description of the phase structure of
YM$_2$ and therefore, the DK solution probably represents the extremum of
the action with respect to $Q$. To our knowledge, this problem has not been
solved so far, and needs more investigations.

Besides this, the chiral version of the large $N$ U($N$) gauge theory on a
two-dimensional sphere of area $A$ has been studied in \cite{CT}.
For small and large areas ($A<A_-$ and $A>A_+$, respectively) a single cut
solution has been found. (That is, a solution for the density function, for
which the region where this function is less than one, is connected.) For
the intermediate region ($A_-<A<A_+$) a two cut solution has been obtained.
(That is, a solution for the density function, for
which the region where this funcion is less than one, consists of two
disconnected parts.) This again, leads to a one parameter family of
solutions for the density function.
Although, as noted in \cite{CT}, the numerical evidence indicates that there
is indeed only one solution in the intermediate region.
We prove this analytically.

In both cases, the number of equations the authors obtain, is one less than
the number of parameters required to determine the density function. In this
paper, we show that there is a missed equation in both cases. This
extra equation removes the arbitrariness of the solutions. For the MP case,
this extra equation fixes $Q$ to be zero. This (apparent) insufficiency of
the number of equations occurs whenever multi-cut solutions are considered
\cite{AKA}.

The partition function of a YM$_2$ theory on a two dimensional compact
Riemann surface of area $A$ and genus $g$ is \cite{Ru2}.
\be \label{1}
{Z}=\s_r (d_r)^{2-2g} \exp \left[-{A\over 2} C_2(r)\right] ,
\ee
where $C_2(r)$ is the second Casimir of the irreducible representation $r$
of the gauge group, and $d_r$ is its dimension. For the gauge group U($N$),
the above summation goes over the lengths of the rows of the Young tableau
($\{ n_1, n_2,\cdots ,n_N\}$) corresponding to the representation $r$. These
should satisfy
\be \label{2}
n_1 \geq n_2\geq \cdots \geq n_N.
\ee
We also have
\bea \label{3}
C_2(r)&=&\s_{i=1}^N[(n_i +N-i)^2-(N-i)^2],\cr
d_r&=&\prod_{1\leq i<j\leq N}(1+{n_i-n_j\over j-i}).
\eea
In the large $N$ limit, one can introduce the continuum variables $x$ and
$n(x)$ [6]:
\be
x:={i\over N}\qquad  n(x):={n_i\over N}.
\ee
Using a change of variable
\be
\p (x):=-n(x)+x-1,
\ee
the partition function takes a simpler form:
\be
 Z=\int \D \p (x)e^{S[\p (x)]},
\ee
where
\be\label{7}
 S[\p (x)]=N^2\left[\int_0^1 \d x\int_0^1 \d y\;\log |\p (x)-\p (y)|
 -{A\over 2}\int_0^1 \d x\;\p^2(x)\right],
\ee
apart from an unimportant constant. The inequality (\ref{2}), expressed
in terms of $\p$, is
\be \label{8}
{\p (x)-\p (y)\over x-y}\geq 1.
\ee
One can introduce the variable $\r [\p (x)]$ as
\be
\r [\p (x)]:={\d x \over \d \p }.
\ee
Then
\be \label{10}
\int_{d=\p (0)}^{a=\p (1)}\r (z)\;\d z=1,
\ee
and, form (\ref{8}),
\be \label{11}
\r (z)\leq 1.
\ee
Forgetting about the constraint (\ref{11}), the saddle point equation
for the action (\ref{7}) is
\be
{Az\over 2}=\P\int_{d}^a {\d \l\;\r (\l)\over z-\l},
\ee
where P indicates the principal value of the integral. The solution of
the above integral equation, along with (\ref{10}), is the well known
semicircle law [6]:
\be  \label{13}
\r (z)={A\over{2\pi}}\sq{{4\over A}-z^2},
\ee
with
\be
a=-d=\sqrt{{4\over A}}.
\ee
The free energy is $F(A)=-{1\over N^2}\log Z$, and its derivative with
respect to the area $A$ is
\be
F'(A)={1\over 2}\int_0^1\d x\;\p^2(x)=\int_{-a}^a\d z\; z^2 \r (z)
     ={1\over 2A}.
\ee
But the equation (\ref{13}) is valid only if $\r (z)\leq 1$, and this holds
for $A\leq A_c=\pi^2$. For $A>A_c$, there is a region where $\r (z)>1$,
and one should search for another solution, which does not violate
(\ref{11}).

DK have considered a symmetric ansatz for $\r (z)$. They take
$\r (z) $ to be equal to one in the region [-$b$, $b$], and less than one in
the region $(-a, -b)\bigcup (b, a)$, where $ 0<b<a$. They obtain the
derivative of the free energy and show that the theory has a third order
phase transition. MP have considered a more general asymmetric ansatz
in which $\r (z)$ is equal to one in the region $[c,b]$, and less than one
in the region $L:=(d, c)\bigcup (b, a)$, where $d<c<b<a$. They obtain three
equations, whereas there are four parameters. They claim that a one
parameter family of solutions exists, which is expressed in terms of $Q$,
the U(1) charge sector of U(N) theory. We show that an equation is missed,
and the solution is unique.

To see this, let's go back to the action (\ref{7}), write the saddle point
equation with respect to $\p (x)$ for $A>A_c$, and express the result in
terms of $\r (z)$. We obtain
\be \label{15}
{Az\over 2}=\P\int_d^a {\d \l\; \r_s (\l )\over z-\l }+
\log\vert {z-c\over z-b}\vert,
\ee
where
\be
\r_s (z)=\cases{{\tilde \r}_s (z),& $z\in L$ \cr
1,&$ z\in [c, b]$\cr }.
\ee
Now define the functions
\bea \label{17}
H_s(z)&:=&\int_d^a {\d \l\; \r_s (\l )\over z-\l },\cr
{\tilde H}_s(z)&:=&\int_L {\d \l\; \tilde\r_s (\l )\over z-\l }.
\eea
The function $H_s(z)$ has a cut  $[d , a]$ and,
$$H_s(z\pm i \e )=\mp i \pi \r _s(z)+\ ({\rm some \ continuous\ function}).
\qquad z\in [d,a]$$
${\tilde H}_s(z)$ has two cuts $[d,c]$ and $[b,a]$. The solution
for ${\tilde H}_s(z)$, and then $H_s(z)$ is \cite{GP},
\be
H_s(z)=\log {z-c\over z-b}+{\sq{(z-a)(z-b)(z-c)(z-d)}\over 2\pi i}
\oint_{ c_L}{A{\l\over 2}-\log{\l -c\over \l-b}\over (z-\l )
\sq{(\l-a)(\l-b)(\l-c)(\l-d)}}\d \l
\ee
where $c_L$ is the contour encircling the cuts $[d,c] $ and $[b,a]$,
leaving the point $z$ out.
One can deform the contour $c_L$  to a contour consisting of three parts:
$ c_{L'}$, which is a contour encircling $[c,b]$; $c_z$, a contour
encircling the pole $\l =z$; and $ c_{\infty}$, a contour at the infinity.
In this way one arrives at
\be \label{21}
H_s(z)= {Az\over 2}-\sq{(z-a)(z-b)(z-c)(z-d)}
\int_c^b{\d \l \over (z-\l )\sq{(a-\l)(b-\l)(\l-c)(\l-d)}}.
\ee
Inserting the above form of $H_s(z)$ in (\ref{17}), and expanding both
sides of (\ref{17}) for large $z$, one can obtain the following equations
from the coefficients of $ z$, $z^0$, and  $z^{-1}$, respectively.
\be \label{22}
{A\over 2}-
\int_c^b{\d \l \over \sq{(a-\l)(b-\l)(\l-c)(\l-d)}}=0,
\ee
\be \label{23}
{a+b+c+d\over 2}
\int_c^b{\d \l \over \sq{(a-\l)(b-\l)(\l-c)(\l-d)}}-
\int_c^b{\l \d \l \over \sq{(a-\l)(b-\l)(\l-c)(\l-d)}}=0,
\ee
\pagebreak
$$
\{-{a(b+c+d)+b(c+d)+cd\over 4}+{a^2+b^2+c^2+d^2\over 8}\}
\int_c^b{\d \l \over \sq{(a-\l)(b-\l)(\l-c)(\l-d)}}
$$
$$
+{a+b+c+d\over 2}\int_c^b{\l \d \l \over \sq{(a-\l)(b-\l)(\l-c)(\l-d)}}-
\int_c^b{\l^2 \d \l \over \sq{(a-\l)(b-\l)(\l-c)(\l-d)}}
$$
\be \label{24}
\qquad \qquad \qquad \qquad \qquad
=\int_d^a \r (\l )\d \l =1.
\ee
The coefficient of $z^{-2}$ of the right hand side of (\ref{17}) is
$\int \l \r_s (\l )\d \l $, which is the charge defined in \cite{MP}.
It is obvious that the three equations (\ref{22}-\ref{24}) are not adequate
for determining $(a,b,c,d)$. MP used the charge $Q$ to obtain the remained
parameter. As we claimed, there exists another equation which must be
considered.
Consider the action (\ref{7}) with the constraint (\ref{10}),
in term of $\r_s (z)$
\be
{\tilde S}=-N^2\Big\{ -\int_d^a\d z\int_d^a \d w \r_s (z) \r_s (w)\log (z-w)
+{A\over 2}\int_d^az^2 \r_s (z) \d z +\l\big[\int_d^a \r_s (z) \d z-1\big]
\Big\},
\ee
where $\l$ is a Lagrange multiplier. Variating ${\tilde S}$ with respect to
$\r_s (z)$ leads to
\be \label{26}
-{1\over{N^2}}{\delta {\tilde S}\over \delta \r_s (z)}=-2\int_d^a \d w \r_s
(w) \log (z-w)+{Az^2\over 2} +\l.
\ee
We must impose
\be \label{127}
{\delta {\tilde S}\over \delta \r_s (z)}=0,\qquad z\in L.
\ee
Differentiating (\ref{127}) with respect to $z$ gives
\be\label{28}
\P\int_d^a {\d w \r (w)\over z-w} ={Az\over 2},\qquad z\in L
\ee
which is the same as (12). But (\ref{127}) has more information than
(\ref{28}). In fact, from (\ref{127})
we also have
\be\label{128}
2\int_d^a \d w  \r_s (w)\log {b-w\over c-w}+{A\over 2}(b^2-c^2)=0,
\ee
which is the difference of eq. (\ref{127}) at $z=b$ and $z=c$. Eq.
(\ref{128}) can be written as
\be \label{27}
\int_c^b \d z\big[\P\int_d^a {\d w \r (w) \over z-w} -{Az\over 2} \big]=0,
\ee
or
\be
\P\int_c^b \d z \int_c^b{\d w {\sq{(a-z)(b-z)(z-c)(z-d)}} \over
(z-w)\sq{(a-w)(b-w)(w-c)(w-d)}}=0.
\ee
This can be written as,
\be \label{31}
\P\int_c^b \d z \int_c^b{\d w \over z-w}+
\int_c^b \d z \int_c^b\d w {{\sq{(z-a)(z-b)(z-c)(z-d) \over
(w-a)(w-b)(w-c)(w-d)}}-1\over z-w}=0.
\ee
The first term is equal to zero. Now, we define the parameters
${\bar x}:={b+c\over 2}$ and $\xi =:{b-c\over 2}$,
and expand (\ref{31}) in terms of $\xi$ (note that $\xi=0$ at $A=A_c$).
This yields
\be
{\pi^2 \xi^2\over 4}{(a+d-2{\bar x})\over (a-{\bar x})({\bar x}-d)}
\big[ 1+\xi^2 {8(\bar x-d)^2+8(a-\bar x)^2+3(a-d)^2\over 32(a-\bar x)^2
(\bar x-d)^2}+\cdots \big]=0.
\ee
The term in the brackets is obviously positive.
So $a+d-2\bar x$ should be zero, which means that,
around the critical area, $a$ and $d$ are symmetric with respect to
$\bar x$. This is the fourth equation, and now, the parameters $a$, $b$,
$c$, and $d$ can be determined uniquely.

There is also a simple and somehow straightforward way to deduce $Q=0$.
Consider the action (\ref{7}), shifting the solution $\p$ by a constant $c$
results
\be
S(c)=-N^2\Big\{ -\int_0^1 \d x\int_0^1 \d y \log \vert \p (x)-\p (y)\vert
+{A\over 2}\int_0^1\d x (\p +c)^2\Big\} .
\ee
Now, as $\p$ is an extermum of $S$, the derivative of $S$ with respect to
$c$ if $\p$ should be zero:
\be
{\d { S}\over \d c}{\big\vert_{c=0}}=0.
\ee
So,
\be
\int_0^1 \p (x) \d x=0.
\ee
This is nothing but the U(1) charge $Q$:
\be
Q=\int z \r (z) \d z=\int_0^1 \p (x)\d x=0.
\ee
MP has shown that this theory is equivalent to a system of $N$
nonrelativistic fermions living on a circle and at the critical area $A_c$
fermion condensation occours \cite{MP}. The partition function for the
fermionic theory is
\be\label{37}
Z=\s_{p_i}\prod_{i>j}(p_i-p_j)^2\exp(-{LT\over 2}\s p_i^2),
\ee
where $p_i$ is the momentum of the $i$'th fermion, and $L$ and $T$ are the
paremeters of the two dimensional space-time cylinder on which the electrons
live. Comparing (\ref{37}) with (\ref{1}), where $C_2(r)$ and $d_r$ are
given by (\ref{3}), it is seen that the U(1) charge $Q$ is $(\s p_i)/N^2$,
or the center of mass momentum divided by $N^2$. The above reasoning leading
to $Q=0$ holds true for $(\s p_i)/N^2=0$.

As a second example, let's consider the chiral YM$_2$ theory. The chiral
U($N$) gauge theory at large $N$ has been studied in \cite{CT}.
For small and large areas ($A<A_-$ and $A>A_+$, respectively) a single cut
solution has been found.
In the intermediate region $A_-<A<A_+$, the number of equations are
insufficient: there are three equations while there are four unknown
parameters. As in the previous case, there is also a missed equation. As
discussed in \cite{CT}, consider the ansatz
\be
\r (z)=\cases{1,&$ z\in R:=[c, b]\bigcup [a, 1/2]$\cr
              {\tilde\r} (z),&$ z\in [d, c]\bigcup [b, a]$.\cr}
\ee
The integral equation for the saddle point is
\be
{Az\over 2}+\log {(z-{1\over 2})(z-b)\over (z-a)(z-c)}=
\P\int_R\d \l {\r (\l )\over z-\l}.
\ee
Using the same argument leading to (\ref{27}), one obtains
\be
\int_c^b\d z[{Az\over 2}+\log {(z-{1\over 2})(z-b)\over (z-a)(z-c)}-
\P\int_R\d \l {\r (\l )\over z-\l}]=0.
\ee
This equation, added to the three equations obtained in \cite{CT}, gives a
complete set of equations. So the solution is, again, unique.

{\bf Acknowledgement} M. Alimohammadi would like to thank the research
vice-chancellor of Tehran University, this work was partially supported by
them.

\end{document}